\documentclass[aps,twocolumn,groupedaddress,showpacs]{revtex4-1}
\usepackage{epstopdf}
\usepackage{physics}
\usepackage{tikz,lipsum,lmodern}
\usepackage[export]{adjustbox}
\usepackage{subcaption}
\usepackage{amsmath}
\usepackage{amssymb}

\usepackage[unicode=true,pdfusetitle,
 bookmarks=true,bookmarksnumbered=false,bookmarksopen=false,
 breaklinks=false,pdfborder={0 0 0},backref=false,colorlinks=true,citecolor=blue,urlcolor=violet 
]{hyperref}
\usepackage{xcolor}
\usepackage{graphicx}
\newcommand{\orcid}[1]{\href{https://orcid.org/#1}{\includegraphics[width=10pt]{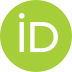}}}
\begin{document}
\title{Characterization of Two-Particle Interference by Complementarity}

\author{Neha Pathania\orcid{0000-0002-3385-4761}}
\email{neha@ctp-jamia.res.in}
\author{Tabish Qureshi\orcid{0000-0002-8452-1078}}
\email{tabish@ctp-jamia.res.in}

\affiliation{Centre for Theoretical Physics, Jamia Millia Islamia, New Delhi-110025, India}

\begin{abstract}
Bohr's Complementarity Principle is quantitatively formulated in terms of
the distinguishability of various paths a quanton can take, and the measure
of the interference it produces. This phenomenon results from the
interference of single-quanton amplitudes for various paths. The
distinguishability of paths puts a
bound on the sharpness of the interference the quanton can produce. However
there exist other kinds of quantum phenomena where interference of
\emph{two-particle} amplitudes results in a two-particle interference,
if the particles are indistinguishable. The Hong-Ou-Mandel (HOM) effect and the
Hanbury-Brown-Twiss (HBT) effect are two well known examples. However, 
two-particle interference is not as easy to define as its single particle
counterpart, and the realization that it involves interference of two-particle
amplitudes, came much later. In this 
work, a duality relation, between the particle distinguishability 
and the visibility of two-particle interference, is derived.
The distinguishability of the two particles, arising from some internal degree
of freedom, puts a bound on the sharpness of the two-particle interference
they can produce, in a HOM or HBT kind of experiment. It is argued that the
existence of this kind of complementarity can be used to characterize
two-particle interference, which in turn leads one to the conclusion that
the HOM and the HBT effects are equivalent in essence, and may be treated
as a single two-particle interference phenonmenon.
\end{abstract}

\maketitle

\section{Introduction}
It is well known that a single particle, better referred to as a quanton,
passing through multiple paths, can interfere with itself, producing an
interference pattern. Bohr's complementarity principle \cite{bohr} can then be
\emph{quantitatively} formulated as a duality relation \cite{cd15}
$\mathcal{D}_{Q} + \mathcal{C} \le  1$ where $\mathcal{D}_{Q}$ is the
path-distinguishability, and $\mathcal{C}$ is the quantum coherence
\cite{plenio} of the quanton. The path-distinguishability is defined in
terms of unambiguous quantum state discrimination (UQSD) \cite{uqsd1,uqsd2}.
Coherence
of the quanton can be measured in a multipath interference experiment
in various ways \cite{tania,tq1,cohrev}. If the path distinguishability
is defined in a different way, one gets a different form of the duality
relation \cite{nduality} $\mathcal{D}^{2} + \mathcal{C}^{2} \le  1$,
where $\mathcal{D}^2=\mathcal{D}_{Q}(2-\mathcal{D}_{Q})$.
However, the two duality relations are the same in essence.
For the special case of only two paths, this relation reduces to the
well known duality relation \cite{englert}
$\mathcal{D}^{2} + \mathcal{V}^{2} \le  1$, where $\mathcal{V}$ is the
conventional visibility of interference, defined as
$\mathcal{V} = \frac{I_{max} - I_{min}}{I_{max} + I_{min}}$,
$I_{max}, I_{min}$ being the maximum and minimum intensity of interference
in a particular region of the interference pattern, respectively.

Apart from this phenomenon which results from interference of single
particle amplitudes, later experiments showed that interference of
two-particle amplitudes is also possible. Such effects have been generically
called two-particle interference. However, it is not always easy to
characterize two-particle interference, and the understanding that it came
from interference of two-particle amplitudes came much later.
Two well known effects that capture the phenomenon well, are the
Hong-Ou-Mandel (HOM) effect \cite{hom,homrev,agata} and the Hanbury-Brown-Twiss (HBT)
effect \cite{hbt,fano,hbtlight1,hbtlight2,tqhbt}. Two-particle interference has
been much studied as well as much debated.
Although for identical particles, the
 two-particle
symmetric and antisymmetric states are entangled, it is not straightforward
 to
identify two-particle interference with this entanglement \cite{franco}. Different
 from
other two-particle quantum effects, it does not seem to originate from the
entanglement between the two particles, and rather appears to have its roots
 in
the fundamental indistinguishability of identical quantum particles.
The HBT effect has also been demonstrated
with massive particles, both of bosonic \cite{hbtatom1,hbtatom2,hbtatom3}
and fermionic \cite{hbtelectrons} nature.
In an interesting development, it was demonstrated that in
a HOM experiment, if the two particles coming from two different paths,
are made partially distinguishable,
it results in the loss of visibility of the HOM interference dip
\cite{2eraser}. If the two particles are fully distinguishable,
the HOM dip completely disappears. Not only that, 
one can set up a ``quantum eraser" in HOM experiment
and recover the HOM dip with maximum visibility \cite{2eraser,herbut}.
In another interesting experiment, a delayed-choice quantum eraser was
demonstrated using thermal light \cite{peng}. In this experiment too
the interference was a two-photon interference.
Although such an effect has not been explored in the HBT
experiment, to our knowledge, we will show that it should exist in the HBT
experiment too. 
These experiments point towards a complementarity involving particle
distinguishability and the visibility of two-particle interference.
We will show here that this complementarity can be \emph{quantified} in the
same way as Bohr's complementarity was quantified by the wave-particle
duality relations. This complementarity will then characterize the two effects
as a unified two-particle inteference phenomenon.
Quantifying complementarity in two-particle interference, and using it to
characterize the two-particle interference,
is the subject of this investigation.

\section{The Hong-Ou-Mandel effect}
\begin{figure}[h]
\centerline{\resizebox{8.5cm}{!}{\includegraphics{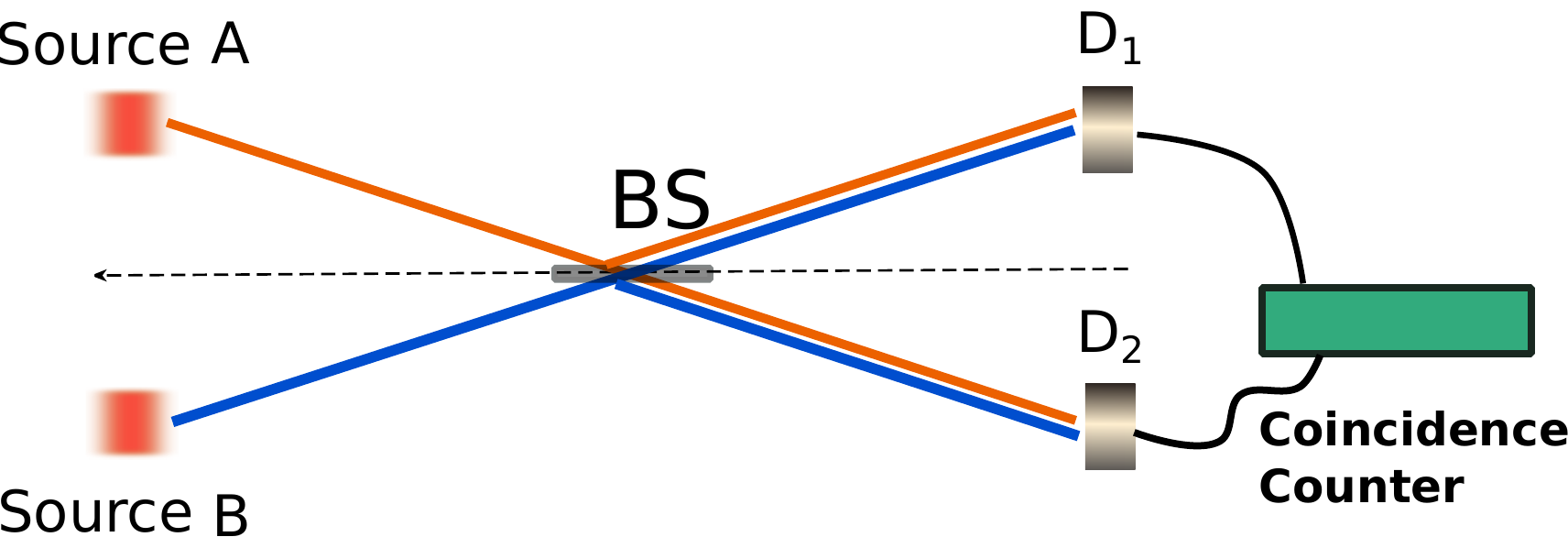}}}
\caption{A schematic diagram for the Hong-Ou-Mandel experiment.
Independent particles from sources A and B meet at the beam-splitter BS,
and then arrive at the detectors $D_1$ and $D_2$.}
\label{HOMexpt}
\end{figure}

We briefly introduce the HOM experiment. Two identical particles emerge
from two spatially separated sources A and B, in the states $|\psi_A\rangle$
and $|\psi_B\rangle$ (see Fig. \ref{HOMexpt}). Assuming that the particles
are bosons, the two-particle state is the following symmetrized state
\begin{equation}
|\Psi_0\rangle = \tfrac{1}{\sqrt{2}}(|\psi_A\rangle_1 |\psi_A\rangle_2 + |\psi_A\rangle_2 |\psi_A\rangle_1),
\end{equation}
where the labels 1,2 are particle labels. The two particles are split by
the 50-50 beam-splitter BS, and reach the \emph{fixed} detectors $D_1,D_2$.
Now let us assume that there exists another
degree of freedom by which the particles from the two
sources can be distinguished. This degree of freedom belongs to the
particles e.g. polarization in the case of photons, and spin in the case of
neutrons.  The combined state of the two particles,
and the additional degree of freedom, can be written as
\begin{equation}
|\Psi\rangle = \tfrac{1}{\sqrt{2}}(|\psi_A\rangle_1 |\psi_B\rangle_2
|d_A\rangle_1|d_B\rangle_2
+ |\psi_A\rangle_2 |\psi_B\rangle_1|d_A\rangle_2|d_B\rangle_1) ,
\end{equation}
where $|d_A\rangle, |d_B\rangle$ are two states of the additional degree
of freedom, which are assumed to be normalized, but not necessarily
orthogonal. If one could distinguish between the state $|d_A\rangle,
|d_B\rangle$, finding a state (say) $|d_A\rangle$ would tell one that the
particular particle came from source A. Since there are two identical
particles involved, the additional degree of freedom has to be
attached to the particle itself. It should be emphasized here that one
cannot have an external path marking device, like that used in single
particle which-way experiments, because when the particles move away from
the path-marker, there is no way to tell which of the two identical
particles is correlated to the path marker. That is
the reason why, for two-particle interference, it does not make sense to
talk of path distinguishability. One can only distinguish between the 
two particles based on some degree of freedom of the particle.
The connection between particle distinguishability and coherence of
the two-particle system has been explored before \cite{Castellini}.

The effect of the beam-splitter on the two states is as follows:
$U|\psi_A\rangle =  \tfrac{1}{\sqrt{2}}(|D_1\rangle-|D_2\rangle)$ and
$U|\psi_B\rangle =  \tfrac{1}{\sqrt{2}}(|D_1\rangle+|D_2\rangle)$,
where $|D_1\rangle,|D_2\rangle$ are the states of a particle at the
detectors $D_1$ and $|D_2$, respectively.
Using this, the two-particle state, after the particles pass through the
beam-splitter, can be written as
\begin{eqnarray}
U|\Psi\rangle &=&  \tfrac{1}{2\sqrt{2}}(|D_1\rangle_1-|D_2\rangle_1)
(|D_1\rangle_2+|D_2\rangle_2)|d_A\rangle_1|d_B\rangle_2 \nonumber\\
&&+  \tfrac{1}{2\sqrt{2}}(|D_1\rangle_2-|D_2\rangle_2)
(|D_1\rangle_1+|D_2\rangle_1)|d_A\rangle_2|d_B\rangle_1 \nonumber\\
\end{eqnarray}
Probability of a coincident count is given by
\begin{eqnarray}
P_C &=& |_1\langle D_1|_2\langle D_2|U|\Psi\rangle|^2
 + |_2\langle D_1|_1\langle D_2|U|\Psi\rangle|^2 \nonumber\\
 &=& \tfrac{1}{4}|(|d_A\rangle_1|d_B\rangle_2 - |d_A\rangle_2|d_B\rangle_1|^2 \nonumber\\
 &=& \tfrac{1}{2}(1-|\langle d_A|d_B\rangle_1||\langle d_A|d_B\rangle_2|) \nonumber\\
 &=& \tfrac{1}{2}(1-|\langle d_A|d_B\rangle|^2)
\end{eqnarray}
This is the minimum intensity of the two-particle interference, as
we have already assumed that the two particles arrive at the beam-splitter
at the same time. If the two particles don't arrive at the beam-splitter
together, each particle acts independently, and is equally likely to land
at $D_1$ or $D_2$. Consequently, the probability of coincident count is
$1/2$. That is the maximum intensity of coincident count, and one can say
$C_{max}=1/2$. The minimum coincident count intensity is given by
$C_{min}=\tfrac{1}{2}(1-|\langle d_A|d_B\rangle|^2)$. The conventional
definition of HOM interference visibility \cite{homrev}, yields
\begin{equation}
 \mathcal{V} = \frac{C_{max} - C_{min}}{C_{max}}
= |\langle d_A|d_B\rangle|^2 .
\label{Vhom}
\end{equation}
At any of the
two detectors, suppose one wants to find out whether a particle came from
source A or source B, the way to do it is to look at the other degree of
freedom of the particle. If one can tell if the state of the particle is
$|d_A\rangle$, it means the particle came from source A, if the state turns
out to be $|d_B\rangle$, it means the particle came from source B. So
the problem of distinguishing the two particles boils down to distinguishing
between the quantum states $|d_A\rangle$ and $|d_B\rangle$. Since the two
states may not be orthogonal, one can use UQSD to unambiguously distinguishing
between them. The optimal probability of successfully distinguishing between
them is given by
$\mathcal{D}_Q = 1 - |\langle d_A|d_B\rangle|$ \cite{cd15}.
Here it is more convenient to define the particle distinguishability as
$\mathcal{D} = \mathcal{D}_Q(2-\mathcal{D}_Q)$ which leads to
\begin{equation}
 \mathcal{D} = 1 - |\langle d_A|d_B\rangle|^2,
\label{D}
\end{equation}
which also happens to be the square of the distinguishability coming from
minimum error discrimination of the two states \cite{englert}. Combining
(\ref{Vhom}) and (\ref{D}), one arrives at
\begin{equation}
 \mathcal{D} + \mathcal{V} = 1 .
\label{dualityhom}
\end{equation}
This is a duality relation involving particle distinguishability $\mathcal{D}$
and the visibility $\mathcal{V}$ of HOM interference. If the states 
$|d_A\rangle$ and $|d_B\rangle$ are orthogonal, the two particle, coming
from different sources, become distinguishable. Consequently they should not
show any HOM effect. Indeed, in such a case the HOM dip disappears and there
is no HOM interference. If $|d_A\rangle$ and $|d_B\rangle$ have partial
overlap, the two particles are partially distinguishable. In that case
the HOM is present, but partially suppressed. Thus the relation 
(\ref{dualityhom}) quantifies the complementarity between particle
distinguishability and two-particle interference.

It is straightforward to see that when $|d_A\rangle,|d_B\rangle$ are 
orthogonal, a quantum eraser can be set up by selecting both particles in 
a state of the internal degree of freedom which has equal overlap with
both $|d_A\rangle$ and $|d_B\rangle$.

\section{The Hanbury Brown-Twiss effect}
\begin{figure}[h]
\centerline{\resizebox{8.0cm}{!}{\includegraphics{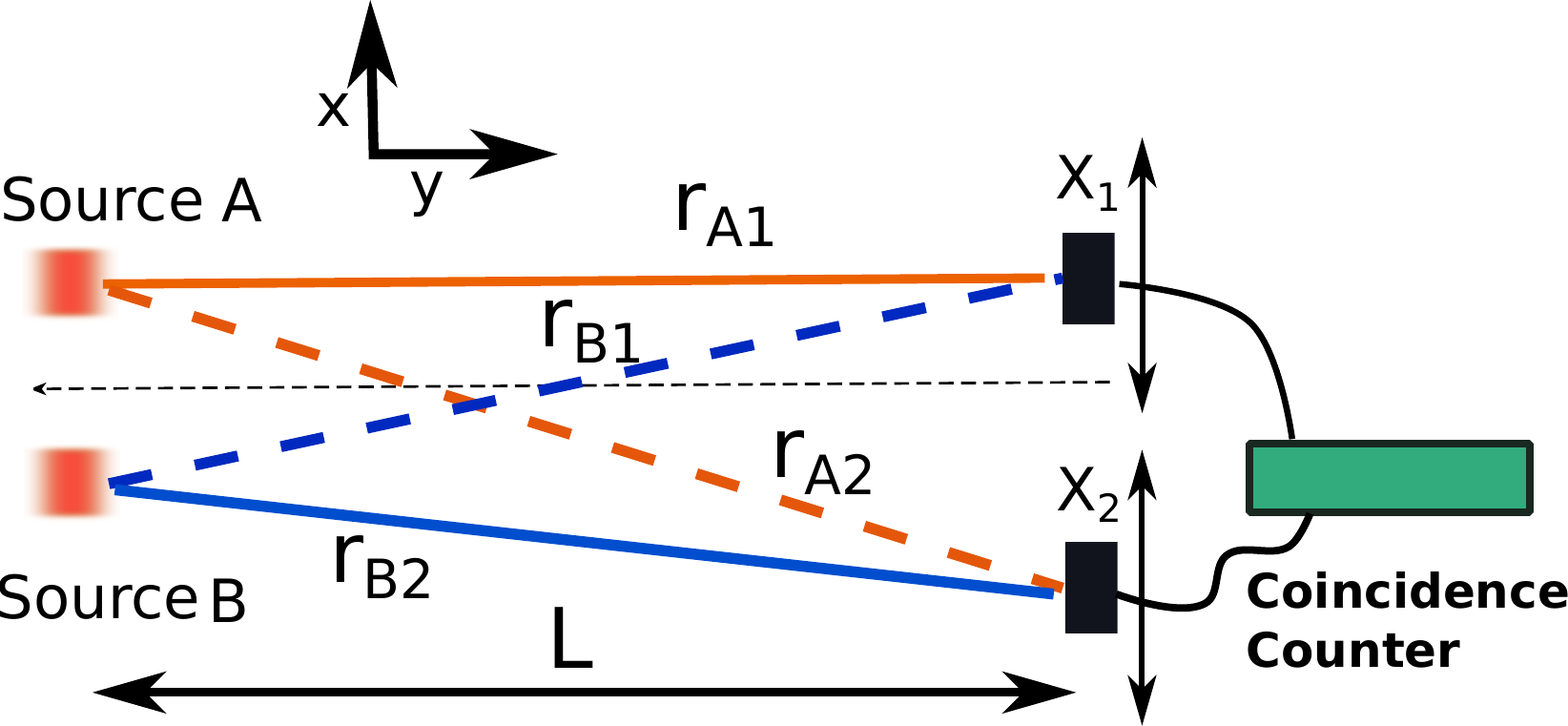}}}
\caption{A schematic diagram for the Hanbury Brown-Twiss experiment.
Independent particles from sources A and B travel
and arrive at the two detectors at $x_1$ and $x_2$.}
\label{HBTexpt}
\end{figure}

The HBT effect was discovered much before the HOM effect, in classical 
radio waves. Later it was demonstrated in classical light \cite{hbt}. Its
applicability and meaning in quantum domain was widely debated and 
misunderstood. The physical understanding of HBT in the quantum domain was
provided by Fano \cite{fano}. In fact, an early two-photon experiment
\cite{ghoshmandel} was believed to demonstrate non-local quantum 
correlations, but was later shown to be just the HBT effect \cite{tqhbt}.
In the HBT experiment two particles emerge from two spatially separated 
sources A and B, and travel to separate, movable detectors at positions
$x_1$ and $x_2$ (see Fig. \ref{HBTexpt}). In our setup, the particles 
travel as wave-packets along the y-axis and spread in both x and y directions.
For the purpose of the HBT effect, their dynamics only along x-axis is
relevant. In the following we will not consider the motion of the particles
along y-axis explicitly. We just assume that the wave-packets travel with
a uniform velocity along y-axis, and after a fixed time they land up at
the detectors.

Let us also assume that the particles carry another degree of freedom,
which may be spin for massive particle and polarization for photons.
This degree of freedom can potentially make the two particles distinguishable.
In the following we assume that the particle, when they emerge from the
two sources, are Gaussian wavepackets centered at $x_0$ and $-x_0$,
traveling along y-axis.
The widths of the wavepackets is assume to be small, denoted by $\epsilon$.
The additional
degree of freedom of the particles, is assumed to have a 
2-dimensional Hilbert space, and the particles emerging from source
A (B) have a state $|d_A\rangle$ ($|d_B\rangle$).

As in the case of HOM experiment, the full wavefunction of the two particles,
with the additional degree of freedom, when they just emerge from the sources,
can be written as
\begin{eqnarray}
\psi(x_1,x_2,0) &=& 
 {1\over\sqrt{\pi}\epsilon}\left(e^{-(x_1-x_0)^2\over\epsilon^2}
e^{-(x_2+x_0)^2\over\epsilon^2}|d_A\rangle_1|d_B\rangle_2\right.\nonumber\\
&&\left. + \eta e^{-(x_1+x_0)^2\over\epsilon^2}
e^{-(x_2-x_0)^2\over\epsilon^2}|d_A\rangle_2|d_B\rangle_1\right),
\label{psi0}
\end{eqnarray}
where $x_1,x_2$ denote the positions of the particles, and $\eta = \pm 1$.
For bosonic particles, the wavefunction should be
symmetric, and $\eta$ should be 1. For fermions, the two-particle wavefunction
should be antisymmetric, requiring $\eta$ to be $-1$.
The particles travel along the y-axis to the two detectors, and also
spread in the x-direction governed by the free particle Hamiltonian
$H = {p_1^2\over 2m} + {p_2^2\over 2m}$, where $m$ is the mass of the
particles.

\begin{figure}[h]
\centerline{\resizebox{8.5cm}{!}{\includegraphics{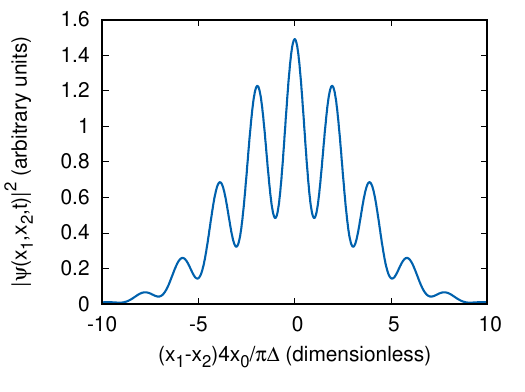}}}
\caption{The probability density of coincident count (\ref{qhbt}), plotted
against detector separation, for $|\langle d_A|d_B\rangle|=0.7$, in the
limit $\Delta \gg \epsilon^2$. The detector separation has been scaled with
the fringe-width, making other parameters unimportant for the interference
pattern.  The interference visibility is noticeably reduced.}
\label{hbtint}
\end{figure}

After a time $t$ the particle reach the detectors.
The amplitude of finding the particles at the detectors at $x_1$ and
$x_2$ then works out to be
\begin{eqnarray}
\psi(x_1,x_2,t) &=& \alpha
\left(e^{-(x_1-x_0)^2\over\epsilon^2+i\Delta}
e^{-(x_2+x_0)^2\over\epsilon^2+i\Delta}|d_A\rangle_1|d_B\rangle_2\right. \nonumber\\
&& \left. + \eta e^{-(x_1+x_0)^2\over\epsilon^2+i\Delta}
e^{-(x_2-x_0)^2\over\epsilon^2+i\Delta}|d_A\rangle_2|d_B\rangle_1\right),
\end{eqnarray}
where $\Delta\equiv 2\hbar t/m$, and $\alpha=\tfrac{1}{\sqrt{\pi(\epsilon+{i\Delta/\epsilon})}}$.
The joint probability density of finding the particles at $x_1$ and $x_2$ is
given by
\begin{eqnarray}
|\psi(x_1,x_2,t)|^2 &=&{2\over \pi\sigma^2}
e^{-2(x_1^2+x_2^2+2x_0^2)\over\sigma^2}
\cosh\big({4(x_1-x_2)x_0\over\sigma^2}\big)\nonumber\\
&&\left(1 + \eta|\langle d_A|d_B\rangle|^2
\frac{\cos\left({4\Delta(x_1-x_2)x_0\over\epsilon^4+\Delta^2}\right)}
{\cosh\big({4(x_1-x_2)x_0\over\sigma^2}\big)}\right),\nonumber\\
\label{qhbt}
\end{eqnarray}
where $\sigma^2 = \epsilon^2+\Delta^2/\epsilon^2$.
The above expression represents a two-particle interference pattern exhibited
in coincident counting of the two detector, as a function of the detector
separation. If the interference is observed sufficiently far from the sources, the situation
is equivalent to the Fraunhofer limit, and $\Delta \gg \epsilon^2$ can be
assumed to hold. In this limit, the above expression simplifies, and yields
a distinct interference pattern (see Fig. \ref{hbtint}). It represents the
HBT effect. The reduced visibility seen in Fig. \ref{hbtint} is due to
$|\langle d_A|d_B\rangle|$ being less than $1$. 
It should be stressed here that
the use of Gaussian wave-packets here is just for the sake of calculational
convenience. A different profile of the wave-packets would lead to the 
same effect. 

Normally one has to be careful
in deciding how to define visibility in a two-particle interference
\cite{jaeger,danko}. However in the HBT setup considered here, it is quite
straightforward, and
one can just use the Michelson fringe contrast for the coincident counts
$\frac{I_{max}-I_{min}}{I_{max}+I_{min}}$. When the wave-packets arrive at
the detectors at $x_1$ and $x_2$, they are expected to be very broad, and
strongly overlapping. The maxima of the intensity will be for the separations
of the two detectors for which the cosine term in (\ref{qhbt}) is equal to
$+1$, whereas the minima will be for separations for which the cosine term is 
equal to $-1$. In the limit $\Delta \gg \epsilon^2$ the cosh term will
be approximately 1. The (ideal) visibility of interference is then given by
\begin{equation}
\mathcal{V} = \frac{I_{max}-I_{min}}{I_{max}+I_{min}}
 = |\langle d_A|d_B\rangle|^2 .
\label{Vhbt}
\end{equation}
In non-ideal conditions the visibility is
$\mathcal{V}\le |\langle d_A|d_B\rangle|^2$. In certain experimental situations
there may be only a few fringes within a rather narrow envelope of the
wavefunction. This will make evaluating the visibility from the interference
pattern more difficult.

\begin{figure}[h]
\centerline{\resizebox{8.5cm}{!}{\includegraphics{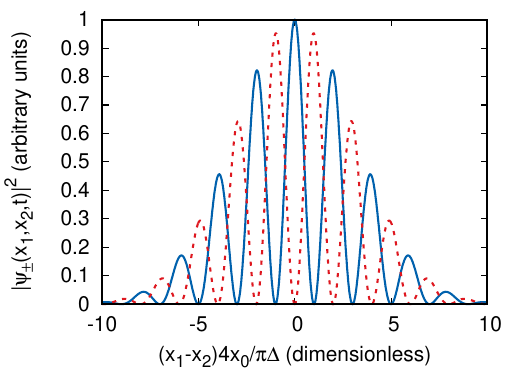}}}
\caption{The probability density of coincident counts (\ref{qhbt}), in
the experimental situation where there are spin filters in front of the
two detectors, depicted by (\ref{erase+}) (solid blue line) and
(\ref{erase-}) (dotted red line), plotted
against detector separation, for $|\langle d_A|d_B\rangle|=0$,
in the limit $\Delta \gg \epsilon^2$.
The two complementary interference patterns have maximum visibility, and
represent quantum erasure.
}
\label{hbteraser}
\end{figure}

As in the HOM experiment, one can distinguish between the two particles 
by analyzing the state of its additional degree of freedom, i.e., the
spin or the polarization. One can then define the particle distinguishability
by (\ref{D}). Using (\ref{D}) and (\ref{Vhbt}), one can write the duality
relation
\begin{equation}
 \mathcal{D} + \mathcal{V} = 1 ,
\label{dualityhbt}
\end{equation}
which is identical to that derived for the HOM effect (\ref{dualityhom}).
Thus it can be thought of as a universal duality relation between
particle distinguishability and two-particle interference.
It should be emphasized here that in the two-particle interference
discussed here, no coherence in the sources is required. Any random 
fluctuation in phases at the source would not affect the interference
and its visibility. These experiments can also be performed with thermal
light \cite{peng}.

From eqn. (\ref{qhbt}) it is obvious that if $|d_A\rangle,|d_B\rangle$ are
orthogonal, no interference will be seen. One can put filters in front of
the two detectors which allow only the particles which have state 
$|d_{\pm}\rangle=(|d_A\rangle\pm|d_B\rangle)/\sqrt{2}$ to pass through.
Identical filter needs to be put in front of both detectors.
In such a situation, the two particle state at the detectors will be
\begin{eqnarray}
\psi_{+}(x_1,x_2,t) &=& \tfrac{\alpha}{2}
\Big(e^{-(x_1-x_0)^2\over\epsilon^2+i\Delta}
e^{-(x_2+x_0)^2\over\epsilon^2+i\Delta} \nonumber\\
&& + \eta e^{-(x_1+x_0)^2\over\epsilon^2+i\Delta}
e^{-(x_2-x_0)^2\over\epsilon^2+i\Delta}\Big),
\label{erase+}
\end{eqnarray}
if both the filters allow $|d_{+}\rangle$. The same state results in
the situation where both the filters allow $|d_{-}\rangle$ state.
This state leads to an interference pattern with maximum visibility, and
corresponds to quantum erasure.
Alternately, one can put a filter which allows only particles with state
$|d_{-}\rangle=(|d_A\rangle-|d_B\rangle)/\sqrt{2}$ in front of detector
at $x_1$, and a filter which allows only particles with state
$|d_{+}\rangle=(|d_A\rangle+|d_B\rangle)/\sqrt{2}$ in front of detector
at $x_2$. In this situation, the two particle state at the detectors will be
\begin{eqnarray}
\psi_{-}(x_1,x_2,t) &=& \tfrac{\alpha}{2}
\Big(e^{-(x_1-x_0)^2\over\epsilon^2+i\Delta}
e^{-(x_2+x_0)^2\over\epsilon^2+i\Delta} \nonumber\\
&& - \eta e^{-(x_1+x_0)^2\over\epsilon^2+i\Delta}
e^{-(x_2-x_0)^2\over\epsilon^2+i\Delta}\Big).
\label{erase-}
\end{eqnarray}
This state also leads to an interference pattern with maximum visibility, but
one which is shifted such that maxima are located at the positions of the
minima of the previous interference pattern (see Fig. {\ref{hbteraser}).
This analysis shows that a quantum eraser is very much possible in an
HBT experiment. Two-particle interference with partially distinguishable
particles is a potentially useful phenomenon. A quantum enhanced microscope
was demonstrated by using two-photon interference and employing the photon
polarization states \cite{Ono}.

\section{Equivalence of HOM and HBT effects}
The HBT effect and the HOM effect have been treated as two distinct effects.
While the HBT effect has also been seen in classical waves, the HOM
effect is believed to be a purely quantum effect. The preceding analysis
of the two effects, and the same kind of complementarity observed in the
two, points to a closer connection between the two. We wish to emphasize
that in our view the two effects are the same. This will be elaborated 
upon in the following discussion. If one considers the single particle
two-slit interference and the single-particle Mach-Zehnder interference
experiment, one may naively think of them as very different experiments.
But a deeper look at the two reveals that they are in essence completely
equivalent. In the two-slit experiment, the particle passes through two
spatially separated slits, and then emerges as two rapidly expanding 
wavepackets which overlap with each other. In different parts of the
overlapping wavepackets, there is constructive or destructive interference.
In a Mach-Zehnder interferometer, a particle is split into two distinct
wavepackets traveling two different paths, which is like the particle
passing through a double-slit. Then the two parts are combined at a 
beam-splitter, and both are split into two parts, so that each beam has
parts from both wavepackets. The essential difference is that the phase
difference between the two paths has to be tuned in such a way that in
one output beam the two packets interfere destructively, and in the other
they interfere constructively. It is as if all the dark fringes of the
double-slit interference are combined into one dark output beam.

Now, the difference between the HBT experiment and HOM experiment is very
similar to that between the two-slit experiment and the Mach-Zehnder
experiment. In the HBT setup, two particles emerge from two spatially
separated sources, and travel in the same direction as expanding 
wavepackets. After some time they overlap, and the joint detection of
the two at different spatial locations, shows a constructive or destructive
interference. For certain separations of the two detectors, there is no
coincident detection. These are the dark fringes. In the HOM setup, like
a Mach-Zehnder setup, two particles emerge from two different source, and
travel two separated well defined paths, and do not overlap. They are
brought together at a beam-splitter, and split into two parts each.
Each of the two beams has parts coming from each particle. Just as in  the
Mach-Zehnder setup the phase difference between the two paths has to be fine
tuned to get null output at one of the two detectors, the time delay 
between the two photons in the HOM experiment has to be fine tuned so that 
the coincident count at the output beams becomes zero. This is equivalent
to coincident count becoming zero for certain separations of the two
detectors in the HBT experiment. Thus the HBT and HOM effects are 
quite analogous to each other, and can be looked upon
as a single two-particle interference phenomenon.

\section{Conclusion}
In conclusion we have analyzed the HOM and HBT effects, and shown that there
exists a quantitative complementarity between the particle distinguishability
and the visibility of the two-particle interference. The more distinguishable
the two particles are, due to some internal degree of freedom, the 
more degraded is the two-particle interference. This complementarity should
be universal in nature and should apply to any two-particle interference.
Such a complementarity points to the fact that there is a single 
phenomenon underlying the HBT and HOM effects.
The quantitative complementarity which is demonstrated here, can
be used to characterize the two particle interference in any variant of
these two experiments.

Neha Pathania acknowledges financial support from DST, India through the
Inspire Fellowship (registration number IF180414).

\end{document}